\documentclass[onecolumn]{IEEEtran}
\usepackage[cmex10]{amsmath}
\usepackage{amssymb}
\usepackage{url}

\newcommand{\ZZ}{{\mathbb Z}}
\newcommand{\FF}{{\mathbb F}}
\newcommand{\ZF}{{\mathbb Z}_4}
\newcommand{\res}{\mbox{Res}}
\newcommand{\card}[1]{\lvert #1 \rvert}
\newtheorem{theorem}{Theorem}
\newtheorem{example}{Example}
\newtheorem{definition}{Definition}
\newtheorem{lemma}{Lemma}

\begin{document}
\title{New upper bounds on binary linear codes \\
and a $\ZF$-code with a better-than-linear Gray image}
\author{Michael Kiermaier, 
	Alfred Wassermann, 
    and Johannes Zwanzger%
\thanks{This work was supported by Deutsche Forschungsgemeinschaft under Grant WA-1666/4. The material of this paper was presented in part at the IEEE Information Theory Workshop Dublin, August 30 -- September 3, 2010.}
\thanks{M. Kiermaier and A. Wassermann are with the Department of Mathematics, University of Bayreuth, D-95440 Bayreuth, Germany}%
\thanks{J. Zwanzger is with Siemens AG, CT RTC ITS SES-DE, Otto-Hahn-Ring 6, 81739 Munich, Germany. 
He was supported by a PhD scholarship from the Studienstiftung des deutschen Volkes (German National Academic Foundation).}
}
\maketitle


\begin{abstract}
Using integer linear programming and table-lookups we prove that there is no binary linear 
$[1988, 12, 992]$ code. As a by-product, the non-existence of binary linear codes with the parameters $[324, 10, 160]$, $[356, 10, 176]$, $[772,11,384]$, and $[836,11,416]$ is shown. 

Our work is motivated by the recent construction of the extended dualized Kerdock code $\hat{\mathcal{K}}^*_{6}$, which is a $\ZF$-linear code having a non-linear binary Gray image with the parameters $(1988,2^{12},992)$.
By our result, the code $\hat{\mathcal{K}}^*_{6}$ can be added to the small list of $\ZF$-codes for which it is known that the Gray image is better than any binary linear code.
\end{abstract}



\section{Introduction}
\IEEEPARstart{I}{n} \cite{kiermaierzwanzger2013} Kiermaier and Zwanzger construct the \emph{extended dualized Kerdock codes} $\hat{\mathcal{K}}^*_{k+1}$ ($k \geq 3$ odd), which are a series of $\ZF$-linear codes with high minimum Lee distance.
The first code $\hat{\mathcal{K}}^*_4$ in this series is a linear $(57,4^4,56)$ code over $\ZF$. Its Gray image is a binary non-linear $(114,2^8,56)$ code.
A table lookup at \cite{Grassl-codetables.de} reveals that the best possible linear code over $\FF_2$ with length $114$ and dimension $8$ has only minimum distance $55$.
That means the minimum distance of the Gray image of this code is higher than the minimum distance of any comparable binary linear code.
For that reason we call the Gray image \textsl{\sl better-than-linear} (\textsl{BTL}).

The second code $\hat{\mathcal{K}}^*_6$ in this series is a linear $(994,4^6,992)$ code over $\ZF$. Its Gray image is a binary non-linear $(1988,2^{12},992)$ code with the Hamming weight enumerator $1 + 4000X^{992} + 31X^{1024} + 64X^{1120}$.
In this note, we prove that this code is BTL, too.
In fact, we show the following result:
\begin{theorem}
If $C$ is a binary linear $[1988, 12, d]$ code, then $d < 992$.
\end{theorem}
As a byproduct we show 
\begin{theorem}
There are no binary linear codes with parameters
$[324, 10, 160]$, $[356, 10, 176]$, $[772,11,384]$, and $[836,11,416]$.
\end{theorem}

For the computer-assisted proof we use a well-known approach using residual codes, table lookups and the MacWilliams equations.
But instead of the usual method to relax the MacWilliams equations and use linear programming to show the non-existence of a code,
we solve the exact MacWilliams equations by using integer linear programming. In order to be able to do this as much weights as possible have
to be excluded beforehand. The use of linear programming has been propagated in~\cite{jaffe1997}, there the split weight enumerator 
has been used. Here, we use the standard weight enumerator of a code. 

\section{$\ZF$-linear codes}
A $\ZF$-linear code $C$ of length $n$ is a submodule of $\ZF^n$. 
The Lee weights of $0$, $1$, $2$, $3\in\ZF$ are $0$, $1$, $2$, $1$, respectively, 
and the Lee weight $w_{\rm Lee}(c)$ of $c\in\ZF^n$ is the sum of the Lee weights of its components.
The Lee distance $d_{\rm Lee}$ of two codewords is defined as the Lee weight of their difference.
The minimum Lee distance $d_{\rm Lee}(C)$ of a $\ZF$-linear code $C$ is defined as 
$d_{\rm Lee}(C) = \min\{w_{\rm Lee}(c)\mid c\in C, c\neq 0\}$ and 
$C$ is called a $(n,\# C, d_{\rm Lee})$ code, where $\# C$ is the number of codewords of $C$.
The Gray map $\psi$ maps $0$, $1$, $2$, $3\in \ZF$ to $(0,0)$, $(1,0)$, $(1,1)$, $(0,1)$, respectively.
It can be extended in the obvious way to a map from $\ZF^n$ to $\FF_2^{2n}$.
The Gray map is an isometry from ($\ZF^n$, $d_{\rm Lee}$) to ($\FF_2^{2n}$, $d_{\rm Ham}$). 
Thus, it maps a $\ZF$-linear $(n, \# C, d)$ code $C$ to an -- in general -- non-linear binary 
$(2n,\#C,d)$ code.

In 
\cite{Hammons-Kumar-Calderbank-Sloane-Sole-IEEETIT40[2]:301-319},
some known BTL codes were found to be Gray images of $\ZF$-linear codes.
Despite many efforts to find more $\ZF$-linear codes with this property,
up to now only a few such examples are known, see Table~\ref{tbl:BTL}.
The column ``lin. bound'' gives the current knowledge on the best possible minimum distance of a comparable binary linear code.
More details can be found in \cite{Kiermaier-2012,kiermaierzwanzger2013}.
In this paper, we add a new example to this list.
\begin{table*}[!t]
\renewcommand{\arraystretch}{1.3}
\caption{$\ZF$-linear codes having a BTL Gray image}
\label{tbl:BTL}
\centering
\begin{tabular}{c|r|c|l}
\hline
& \bfseries Gray image & \bfseries lin. bound & \bfseries $\ZF$-code \\
\hline\hline
& $(14, 2^6, 6)$ & 5& Heptacode (shortened Octacode) \cite{Conway-Sloane-1993}; code $\mathrm{C}(\mathfrak{T}_3)$ for $\mathbb{G} = \ZF$ in \cite{Honold-Landjev-2005}. \\
& $(16, 2^8, 6)$ & 5& Octacode \cite{Conway-Sloane-1993}. Its Gray image is the Nordstrom-Robinson code \cite{Nordstrom-Robinson-1967}. \\
& $(58, 2^7, 28)$ & 27& code $\hat{\mathcal{C}}$ in \cite{Kiermaier-Zwanzger-2011}; lengthened Simplex code $\hat{\mathcal{S}}_{2,3}$ in \cite{Kiermaier-2013}. \\
& $(60, 2^8, 28)$ & 27& doubly shortened $\ZF$-Kerdock code. \\
& $(62, 2^{10}, 28)$ & 26--27& shortened $\ZF$-Kerdock code; code $\mathrm{C}(\mathfrak{T}_5)$ for $\mathbb{G} = \ZF$ in \cite{Honold-Landjev-2005}.  \\
& $(62, 2^{12}, 26)$ & 24--25& punctured $\ZF$-Kerdock code.  \\
& $(64, 2^{11}, 28)$ & 26--27& expurgated $\ZF$-Kerdock code. \\
& $(64, 2^{12}, 28)$ & 25--26& $\ZF$-Kerdock code \cite{Kerdock-1972,Hammons-Kumar-Calderbank-Sloane-Sole-IEEETIT40[2]:301-319}. \\
& $(114, 2^8, 56)$ & 55& extended dualized Kerdock code $\hat{\mathcal{K}}^*_{4}$ \cite{kiermaierzwanzger2013}. \\
& $(372, 2^{10}, 184)$ & $\leq 183$ & dualized Teichm\"{u}ller code $\mathcal{T}^*_{2,5}$ \cite{kiermaierzwanzger2013}, see also \cite{Kiermaier-2012,Kiermaier-2013}. \\
\textbf{new} &  $(1988,2^{12},992)$ & $\mathbf{\leq 991}$ & extended dualized Kerdock code $\hat{\mathcal{K}}^*_{6}$ \cite{kiermaierzwanzger2013}. \\
&  $(2^{k+1}, 2^{2^{k+1} - 2(k+1)}, 6)$ & $\leq 5$ & $\ZF$-Preparata code for all $k\geq 3$ odd \cite{Preparata-1968,Hammons-Kumar-Calderbank-Sloane-Sole-IEEETIT40[2]:301-319,Brouwer:1993}. \\
\hline
\end{tabular}
\end{table*}

In \cite[Th.~5]{kiermaierzwanzger2013} a new series of $\ZF$-linear codes of high minimum Lee distance is given:
\begin{theorem}\label{kiermaierzwangerseries}
    For odd $k \geq 3$, the extended dualized Kerdock code $\hat{\mathcal{K}}^*_{k+1}$ is a $\ZF$-linear code with the parameters
    $$
        (2^{2k}-2^k+2^{(k-3)/2},\; 4^{k+1},\; 2^{2k}-2^k)\, .
    $$
\end{theorem}

\begin{example}
The first two codes in the series of Theorem~\ref{kiermaierzwangerseries} have the following parameters:
    \begin{itemize}
        \item $k=3$: $(57,2^8,56)$ with Gray image $(114,2^8,56)$,
        \item $k=5$: $(994,2^{12},992)$ with Gray image $(1988,2^{12}, 992)$.
    \end{itemize}
\end{example}
The code with parameters $(114,2^8,56)$ is known to be BTL.
In the following, we will show that the $(1988,2^{12}, 992)$ code is BTL, too.

\section{Preliminaries}

\subsection{The MacWilliams equations}
Let $C$ be a binary linear code and 
$A_i$ the number of codewords of weight $i$, $1\leq i \leq n$.
Its weight enumerator is the polynomial
$$
W(C) = \sum_{i=0}^n A_i X^i\; .
$$

\begin{theorem}[MacWilliams equations \cite{bok:MW}]\label{macwilliams}
For $0\leq j\leq n$:
    $$
        \card{C} \cdot A_j^\perp = \sum_{i=0}^n K_j^{n,q}(i)\cdot A_i\;,
    $$
where 
$$ 
	K_k^{n,q}(x) = \sum_{j=0}^k (-1)^j(q-1)^{k-j}\binom{x}{j}\binom{n-x}{k-j}
$$
are the Krawtchouk polynomials.
\end{theorem}
From the MacWilliams equations the Pless power moments can be derived, see e.g. \cite[Ch. 7.3]{Huffman-Pless-2003}.
The first three power moments in the binary case are
\setlength{\arraycolsep}{0.0em}
\begin{eqnarray}
\sum_{j=0}^n A_j &{}={}& 2^k \label{P0}\\
\sum_{j=0}^n jA_j &{}={}& 2^{k-1}(n - A_1^\perp)  \label{P1}\\
\sum_{j=0}^n j^2A_j &{}={}& 2^{k-2}\bigl(n(n+1) - 2nA_1^\perp + 2A_2^\perp\bigr)\,. \label{P2}
\end{eqnarray}
\setlength{\arraycolsep}{5pt}

P. Delsarte~\cite{delsarte1972} uses Theorem \ref{macwilliams} to find new upper bounds 
for code parameters by linear programming.
By setting $x_i:= A_i/\card{C}$ and using the fact the coefficients of weight enumerators are non-negative numbers, the MacWilliams equations 
imply the inequalities
$$0 \leq \sum_{i=0}^n K_j^{n,q}(i)\cdot x_i\;,\quad 0\leq j\leq n$$
with the additional restrictions on $x_i$:
\begin{itemize}
    \item $0\leq x_i\leq 1$,
    \item $x_0=1/\card{C}$,
    \item $x_i=0$, $i=1,\ldots,d-1$,
    \item $\sum_{i=0}^n x_i = 1$.
\end{itemize}
Finding the exact solution of the MacWilliams equations is an integer linear feasibility problem which is a variant of 
the integer linear programming (ILP) problem, see e.g. \cite{Nemhauser:1988}:

Determine
$$A_i, A_j^\perp \in\ZZ \quad (0\leq i,j\leq n)$$
    such that 
    $$
        0 = \card{C}\cdot A_j^\perp - \sum_{i=0}^n K_j^{n,q}(i)\cdot A_i\;
        \quad    \mbox{ for } 0\leq j\leq n
    $$
    and
    \begin{itemize}
        \item $0\leq A_i < \card{C}$, $0\leq A_j^\perp < \card{C^\perp}$,
        \item $A_0= A_0^\perp=1$, 
        \item $\sum_{i=0}^n A_i = \card{C}$, $\sum_{i=0}^n A^\perp_i = \card{C^\perp}$.
    \end{itemize}
For solving ILPs we will use the algorithm \cite{wassermann:02} which is based on lattice point enumeration.

\subsection{Residuals and the Griesmer bound}
\begin{definition}
For a linear $[n,k]$ code $C$ and a codeword $c\in C$
the \textsl{residual code} $\res(C,c)$ of $C$ with respect to $c$ is the code
$C$ punctured on all nonzero coordinates of the codeword $c$.
\end{definition}
In \cite{HenkCA1981197}, a lower bound on the minimum distance of $\res(C,c)$  of a binary code $C$ is given.
This has been generalized to arbitrary prime powers $q$ by \cite{Hill:1992:OTL:142976.142992}.
\begin{theorem}[\cite{Hill:1992:OTL:142976.142992}]\label{griesmerstep}
For a linear $[n,k,d]$ code $C$ over $\FF_q$ and a
codeword $c\in C$ having weight $w<dq/(q-1)$
the residual code $\res(C,c)$ is an $[n-w,k-1,d']$ code with
       $$d'\geq d -w+\lceil w/q\rceil.$$
\end{theorem}

The repeated application of Theorem~\ref{griesmerstep} to codewords $c$ of minimum weight leads to the Griesmer bound, which has been formulated for binary linear codes in \cite{Griesmer-1960} and was generalized to arbitrary $q$ in \cite{Solomon-Stiffler-1965}.
\begin{theorem}[Griesmer bound \cite{Solomon-Stiffler-1965}]\label{griesmer}
For a binary linear $[n,k,d]$ code, we have
\[
	n \geq \sum_{i=0}^{k-1} \left\lceil\frac{d}{2^i}\right\rceil\text{.}
\]
\end{theorem}

\section{Non-existence of a binary linear $[1988,12,992]$ code}
We assume that there exists a binary linear $[1988,12,992]$ code.
\begin{theorem}[\cite{simonis}]\label{simonis}
Any linear code $C\subset \FF_q^n$ of dimension $k$ and minimum weight $d$ can be transformed into a code $C'\subset \FF_q^n$
with the same parameters such that $C'$ possesses a basis of weight $d$ vectors.
\end{theorem}

From Theorem \ref{simonis} 
we get the existence of a binary linear $[1988,12,992]$ code $C$ 
which has a basis consisting of codewords of minimum weight $992$.
As the sum of two binary words of even weight is again of even weight, all the weights of $C$ are even.

\subsection{Table lookup}
Many weights of $C$ can be excluded by applying Theorem~\ref{griesmerstep} iteratively and by table lookup at \cite{Grassl-codetables.de,jaffe-tables}.
    
\begin{example}
Suppose there exists a codeword of weight $1000$ in $C$.
Applying Theorem~\ref{griesmerstep} 
for a codeword of weight $1000$ leads to a  $[988,11,\geq 492]$ code. Now we iteratively 
apply Theorem~\ref{griesmerstep} to codewords of minimum weight and arrive at
a $[496,10,246]$ code and finally at a $[250,9,\geq 123]$ code. 
A table lookup at \cite{Grassl-codetables.de} shows that the upper bound for a binary linear $[250,9]$ code is $122$.
It follows, there is no binary linear $[1988,12,992]$ code having a codeword with weight $1000$.
\end{example}

In the same way all nonzero weights can be excluded except the twelve weights
$992$, $1008$, $1024$, $1056$, $1088$, $1152$, $1216$, $1280$, $1344$, $1984$, $1986$, and $1988$.

\subsection{The weights $\geq 2d$}\label{geq2d}
By using appropriate linear combinations of codewords the weights $1986$ and $1988$ can be excluded,
e.g. addition of the codeword of weight $1988$ and a codeword of minimum weight $992$ would
give a codeword of weight $996$.

Excluding the weight $2d=1984$ requires a little bit more work.
Adding a codeword $c_1$ of weight $1984$ and an arbitrary codeword $c_2$ of weight $992$ might be again a codeword of weight $992$. 
More precisely, $w_{\rm Ham}(c_1 + c_2) \geq 992$ with equality if and only if the support of $c_2$ is contained in the support of $c_1$.
Hence the existence of a codeword $c_1$ of weight $1984$ implies that the supports of all the codewords of minimum weight $992$ are contained in the support of $c_1$.
Since $C$ has a basis of minimum weight words, the four coordinates not in the support of $c_1$ are zero coordinates of $C$, and shortening $C$ in these four coordinates yields a binary linear $[1984,12,992]$ code.
This is a contradiction to the Griesmer bound: The length of a binary linear code of dimension $12$ and minimum distance $992$ is at least
\[
	\sum_{i=0}^{11} \left\lceil \frac{992}{2^i}\right\rceil
	= 1985\text{.}
\]

\subsection{The weight $1344$}\label{subsect:1344}
If $C$ has a codeword of weight $1344$, then the twofold application of Theorem~\ref{griesmerstep} gives a binary linear $[324,10,\geq 160]$ code.
In fact, the parameters are $[324,10,160]$, since a minimum distance $\geq 161$ is impossible by the Griesmer bound.

Again, using \cite[Th.~2.7.8]{Huffman-Pless-2003} we get the existence of an even binary linear $[324,10,160]$.
The application of Theorem \ref{griesmerstep} and table lookups to this parameter set show 
that the only possible nonzero weights of a binary linear $[324,10,160]$ code are 
$160$, $320$, $322$, and $324$.
The weights $\geq 2d = 320$ can be excluded as in Section \ref{geq2d}, using that the length of a binary linear code of dimension $10$ and minimum distance $160$ is at least $322$ by the Griesmer bound.
This leaves $160$ as only possible nonzero weight.

The power moment (\ref{P1}) gives the equation
$$2^{9}\cdot324 - (2^{10}-1)\cdot160 = 2208 = 2^{9}\cdot A_1^\perp$$
in contradiction to $A_1^\perp\in\mathbb Z$.
This shows
\begin{lemma}\label{code324}
A binary linear $[324, 10, 160]$ code does not exist.
\end{lemma}
In particular, the code $C$ does not have codewords of weight $1344$.

\subsection{The weight $1280$}
If $C$ has a codeword of weight $1280$, the strategy of Section~\ref{subsect:1344} leads to the existence of an even binary linear $[356,10,176]$ code.
Table lookup shows that the only possible nonzero weights are
$176$, $192$, $352$, $354$, and $356$.
The weights $\geq 2d = 352$ can be excluded as in Section \ref{geq2d} since the Griesmer bound is equal to $354$.

From (\ref{P0}) it follows that $A_{176} + A_{192} = 2^{10} - 1$. Then, equation (\ref{P1}) gives
$A_{192} = 139 - 32 A_1^\perp$ and $A_{176} = 884 + 32 A_1^\perp$. Using this in equation (\ref{P2}) gives
$$
12A_1^\perp + A_2^\perp = -56\text{,}
$$
which has no solution for nonnegative values of $A_1^\perp$ and $A_2^\perp$.
Therefore, we have 
\begin{lemma}\label{code356}
A binary linear $[356, 10, 176]$ code does not exist.
\end{lemma}

\subsection{The weight $1216$}
If $C$ has a codeword of weight $1216$, we descend to an even $[772,11,384]$ code like in Section~\ref{subsect:1344} .
Application of Theorem \ref{griesmerstep} and table lookups show that the only possible nonzero
weights of a binary linear $[772,11,384]$ code are $384$, $416$, $448$, $768$, $770$, and $772$.
The weights $\geq 2d = 768$ can be excluded as in Section \ref{geq2d} since the Griesmer bound is equal to $769$.

Application of Theorem \ref{griesmerstep} to  $w=416$ and $w=448$ would lead to $[356,10,176]$ and $[324,10,160]$ codes, which do not exist by Lemma~\ref{code324} and~\ref{code356}.
Thus, the only possible remaining weight is $384$.
Using the power moment (\ref{P1}) immediately tells us that such a code does not exist.
So we have:

\begin{lemma}\label{code772}
A binary linear $[772,11,384]$ code does not exist.
\end{lemma}

\subsection{The weight $1152$}
If $C$ has a codeword of weight $416$, we descend to an even $[836,11,416]$ code like in Section~\ref{subsect:1344}.
Application of Theorem \ref{griesmerstep} and table lookups show that the only possible nonzero weights of a binary linear $[836,11,416]$ code are $416$, $448$, $480$, $512$, $832$, $834$, and $836$.
The weights $\geq 2d = 832$ can be excluded as in Section \ref{geq2d} since the Griesmer bound is equal to $834$.

Again, Theorem \ref{griesmerstep} for $w=480$ and $w=512$ would lead to the non-existing   
$[356,10,176]$ and  $[324,10,160]$ codes.   

From (\ref{P0}) it follows that $A_{416} + A_{448} = 2^{11} - 1$. Then, equation (\ref{P1}) gives
$A_{448} = 141 - 32 A_1^\perp$ and $A_{416} = 1906 + 32 A_1^\perp$. Using this in equation (\ref{P2}) gives
$$
28 A_1^\perp + A_2^\perp = -116\text{,}
$$
which has no solution for nonnegative values of $A_1^\perp$ and $A_2^\perp$.
It follows
\begin{lemma}\label{code836}
A binary linear $[836,11,416]$ code does not exist.
\end{lemma}

\subsection{The remaining weights}\label{remainingweight}
At this point the remaining possible nonzero weights of the $[1988,12,992]$ code are
$992$, $1008$, $1024$, $1056$, and $1088$.

Furthermore, we have $A^\perp_1 = 0$:
Otherwise, $C$ has a zero coordinate.
Puncturing in this coordinate yields a binary linear $[1987,12,992]$ code.
After three applications of Theorem~\ref{griesmerstep}, we get the existence of a binary linear $[251,9,\geq 124]$ code in contradiction to the online table \cite{Grassl-codetables.de}.

Therefore, in the ILP there remain the $5$ variables $A_i$ with $i\in \{992, 1008, 1024, 1056, 1088\}$ bounded by $0\leq A_i\leq 4096$ and the $1987$ variables $A_j^\perp$ with $j\in\{2,\ldots,1988\}$ bounded by $0\leq A_j^\perp\leq 2^{1976}$.

Due to the large number of variables and the huge absolute values of the coefficients and bounds, the resulting ILP is still very difficult to solve.
At the time being, standard Integer Program solvers are not able to handle this problem.
However, it turned out to be small enough to be attacked by the specialized method of \cite{wassermann:02}. 
Using the LLL algorithm from the NTL library by V. Shoup~\cite{shoupntl} and our own NTL-implementation of lattice point enumeration we find that the ILP has no solution in about three hours on a standard PC.
    
\bigskip
It follows that a binary linear $[1988,12,992]$ code does not exist.
Consequently, the $(1988,2^{12},992)$ Gray image of the $\ZF$-linear extended dualized Kerdock code $\hat{\mathcal{K}}^*_{6}$ is BTL.

\bigskip
We would like to conclude this note with the following open question: 
Are there any further codes in the series $\hat{\mathcal{K}}^*_{k+1}$ whose Gray image is BTL?

\section*{Acknowledgement}
We thank the anonymous referee for helpful comments.

\bibliographystyle{IEEEtran}


\end{document}